# All-Optical High-speed Programmable Nonlinear Activation Functions using a Fabry-Perot Laser

Mladen Banović[1], Petar Atanasijević[1], Antonios Prapas[2], Christos Pappas[2], Jasna Crnjanski[1], Apostolos Tsakyridis[2], Miltiadis Moralis-Pegios[2], Konstantinos Vyrsokinos[2], Milanka Lović[1], Nina Zdravković[1], Milena Mićić[1], Marko Krstić[1], Slobodan Petričević[1], Nikos Pleros[2], Dejan Gvozdić[1]

*Abstract*—The threads of photonics are eagerly awaited to redefine the future of neuromorphic data processing, especially as the computing-intensive artificial intelligence models become an unavoidable part of our everyday lives. Still, there is much to be improved within the domain of photonic nonlinear activation functions, as the programmable, all-optical, energy-efficient nonlinearities remain beyond the grasp of today's state of the art. In this paper, we address the issue at hand and propose a novel approach in the realization of high-performing all-optical photonic activations. Through simulations and experiments, we show that Fabry-Perot laser diodes (FP-LDs) exhibit richness and high programmability of their nonlinear response to input optical pulses with widths as low as 25 ps. We demonstrate a variety of sigmoid-like and inverted PReLU-like trends to be used as all-optical activation functions in photonic neural networks, testing their performance in stringent, real-life training scenarios with randomized data patterns at repetition rates up to 10 GHz. The programmability of activations is shown using a multitude of experimental operating parameters, among which we highlight the power variation of an additional continuous wave laser, injected into the FP-LD, enriching our approach with all-optical control of all-optical activations. With very low static power consumption of our active element, we achieve a record-breaking energy draw on the order of pJ to hundreds of fJ per nonlinear operation.

## I. INTRODUCTION

THE proliferation of highly demanding artificial intelligence workloads in modern computational systems has fueled a drive towards custom specialized hardware. In this endeavor, photonic solutions have arisen as a promising technological candidate as they can offer high computing rates at a reduced energy envelope. However, they are still technologically challenged to fit the demands of integration, reproducible fabrication, and reconfiguration to achieve high versatility. Neuron's nonlinear activation function (NAF) remains a particularly challenging target. The proposed photonics hardware-based solutions are either opto-electro-optical (OEO) or all-optical (AO), with both however, often requiring electrical control to achieve any degree of reconfiguration.

The electrical stage of the reconfigurable OEO schemes is typically analog [1, 2], where a part of the input optical signal is converted to the analog voltage using a photodiode (PD), with a transimpedance amplifier (TIA) chain. This voltage is subsequently used to modulate a Mach-Zehnder Interferometer (MZI) using a phase modulator, or to self-phase modulate the rest of the signal in a microring resonator (MRR) assisted MZI, yielding a set of up to 15 NAFs. Alternatively, Fard [3] demonstrates arbitrary NAF generation using digital domain reconfiguration. A part of the optical signal is converted to a digital word, while a microcontroller with a lookup table controls a thermal phase shifter within one arm of the MZI. More recently, a programable TIA driven by a balanced PD was experimentally demonstrated to achieve 7 different NAFs including the hyperbolic tangent, at line rates up to 10 Gb/s [4]. This architecture was further tested within the optics-informed training framework, bridging the gap between ideal conditions and liferealistic applications [5]. This leads to the conclusion that OEO schemes provide a reliable and flexible way to achieve a wide variety of NAFs even at high speeds, although at a price of high energy consumption, usually dominated by TIAs and electrical amplifiers.

Contrary to the power-hungry, latency-burdened OEO schemes, the AO paradigm avoids transferring the signals between domains, aiming for faster, more energy-efficient devices. Among the different approaches, nonlinear MRRs

Preprint posted on March 27th 2025.

The research was supported by Science Fund of the Republic of Serbia (#7750121, ORCA-LAB), and by Ministry of Science, Technological Development and Innovation of the Republic of Serbia under contract number: 451-03-137/2025-03/200103. The research was partially conducted in the premises of the Palace of Science, Miodrag Kostić Endowment.

Aristotle University of Thessaloniki also acknowledges support by the EC via H2020 Projects GATEPOST (101120938) and HAETAE (101194393).
*(Corresponding author: Mladen Banović).*

The authors are with the: [1]University of Belgrade - School of Electrical Engineering, 11120 Belgrade, Serbia and [2]Centre for Interdisciplinary Research and Innovation, Informatics Dept. Aristotle University of Thessaloniki, Greece utilizing the thermo-optic effect stand out for their extremely low footprints and passive nature, albeit their speed is severely limited by slow thermal transients [6, 7]. In contrast, MZI schemes utilizing semiconductor optical amplifiers (SOAs) realizing sigmoid-like activations were shown to provide speeds of over 3 Gb/s at the cost of increased power consumption [8]. A compromise is offered by the MZI cavity loaded with MRRs, programmed using microheaters, promising Gb/s speeds at moderate energy consumptions [9, 10]. Finally, the state-of-the-art energy consumption of under 1 pJ per nonlinear operation (NOp) was reached using a periodically poled thin-film lithium niobate nanophotonic waveguide [11]. Rectified, Exponential, and Gaussian Error Linear Unit (ReLU, ELU and GELU), were reported by switching between the effects of frequency doubling and degenerate optical parametric amplification using the phase difference between the signal and the bias input pulses. This unconventional approach comes with its limitations in cascadability due to the necessity for continuous pulse-signal phase control. Furthermore, the operation of this NAF was experimentally shown only for repetition rates significantly under 1 GHz. Unfortunately, a common issue for all the aforementioned AO approaches is the reduced reconfigurability compared to their OEO counterparts.

Injection-locked (IL) LDs have been demonstrated to provide reliable and simple AO NAFs, while consuming relatively low energy and allowing for reconfigurability [12]. However, typical optical injection schemes can hardly exceed GHz bandwidths [13, 14]. In this paper, we expand on our previously demonstrated AO NAFs [15] relying on IL FP-LDs to increase the operational rate by more than threefold and allow for up to 10 Gb/s line-rates while retaining the reconfigurable character in sigmoid- and Parametric ReLU-like (PReLU) NAF families. Furthermore, we demonstrate the improved NAF reconfigurability either by means of chirp-filtering using an optical bandpass filter (BPF) at the LD's output or the injection of an additional continuous wave (CW) optical input, moving from a single optical injection (single-OI) to a dual optical injection (dual-OI) scheme. Alongside experimental demonstrations, we provide thoroughly explained theoretical simulations based on an improved theoretical model. With the additional CW optical input, we break the sub-pJ energy consumption per nonlinear operation barrier, demonstrating adaptive AO NAFs for 25 ps pulses.

## II. MODEL OF THE FP-LD UNDER OPTICAL INJECTION

To describe the LD's dynamics more accurately, we extend the previously used rate equation model [16, 17] by incorporating carrier transport and parasitic effects in the active region. The active region is composed of 3 InGaAsP/InGaAlAs quantum wells and a separate confinement region (SCH), with the cross-section and design-related details following the guidelines provided in [18]. The model governs the dynamics of the carrier density in the barrier (continuum), $n_b$, and in the well states $n_w$, through a system of coupled rate equations in the reservoir framework:

$$\frac{dn_b}{dt} = \eta_{inj}\frac{I}{qV_b} - Q_b(n_b) - \frac{n_b}{\tau_{bw}} + \frac{n_w}{\tau_{wb}}\frac{V_w}{V_b} \quad (1)$$

$$\frac{dn_w}{dt} = \frac{n_b}{\tau_{bw}}\frac{V_b}{V_w} - Q_w(n_w) - \frac{n_w}{\tau_{wb}} - \sum_{j\in J} v_g \frac{g(n_w,\lambda_j)}{1+\epsilon\cdot S_{tot}}S_j \quad (2)$$

Here, $\eta_{inj} = 0.8$ is the injection efficiency, $I$ is the electrical current reaching the MQW region, $v_g$ is the group velocity, $V_w \approx 9\cdot 10^{-12}$ cm$^3$ is the volume of the wells and $V_b \approx 9\cdot 10^{-11}$ cm$^3$ the total volume of the SCH and active region, $\tau_{bw} = 5$ ps and $\tau_{wb} = 50$ ps are capture and escape times to and from the wells, respectively, while $g(n_w,\lambda_j)$ is the carrier-dependent material gain at the wavelength $\lambda_j$ (in vacuum) corresponding to the $j$-th longitudinal mode in FP-LD cavity, $\epsilon = 2.85\cdot 10^{-17}$ cm$^3$ is the nonlinear gain suppression coefficient and $S_{tot} = \sum_j S_j$ is the total photon density for all longitudinal modes. The total recombination rate is calculated as $Q_{w/b} = A_{w/b}n_{w/b} + B_{w/b}n_{w/b}^2 + C_{w/b}n_{w/b}^3$, where index $w/b$ denotes well/barrier region, $A_{w/b} = 1.1\cdot 10^8/1.1\cdot 10^7$ s$^{-1}$ and $C_{w/b} = 4.5\cdot 10^{-28}/3\cdot 10^{-29}$ cm$^6$/s are Shockley–Read–Hall (SRH) and Auger recombination coefficients, respectively, $B_b = 7\cdot 10^{-11}$ cm$^3$/s the bimolecular recombination coefficient in the barrier, while $B_w n_w^2 = R_{sp}(n_w)$ is the spontaneous recombination rate in the wells. The material gain $g(n_w,\lambda_j)$ and spontaneous recombination rate $R_{sp}(n_w)$ dependencies at the signal wavelength are numerically calculated using the 8×8 k.p method [16, 19].

The system comprises an additional 101 equation describing photon density ($S_j$) dynamics for each supported longitudinal mode $j$, including the modes $m \in M = \{p, c\}$ under intermodal optical injection ($S_m$) for which the coupling of photon density and phase of the injected light ($S_m^{inj}$) is introduced through the Kronecker delta $\delta_{jm}$ and the external light coupling coefficient $k_c = 1.34\cdot 10^{11}$ s$^{-1}$:

$$\frac{dS_j}{dt} = A_j S_j + B(n_w) + 2k_c\delta_{jm}\sqrt{S_m S_m^{inj}}\cos\theta_m \quad (3)$$

In (3), $B(n_w) = \Gamma\beta_{sp}R_{sp}(n_w)$ is the effective spontaneous emission, and $A_j = \Gamma v_g g(n,\lambda_j)/(1+\epsilon S_{tot}) - 1/\tau_p$ is the effective rate of the stimulated photon generation where $\tau_p = 2.73$ ps stands for the photon lifetime, $\Gamma = 0.05$ is the optical confinement factor, and $\beta_{sp} = 1.78\cdot 10^{-4}$ is the spontaneous emission coupling factor. In previous equations, $\lambda_j$ represents the wavelength of the unlocked side-mode $j$. For the IL side mode $m$ the wavelength is denoted by $\lambda_m$ and is detuned by $\Delta\lambda_m > 0$ with respect to the unlocked case $(\lambda_m - \Delta\lambda_m)$. Injected/output photon density $S_m^{inj/out}$ is related to the injected/output power $P_m^{inj/out}$ via $S_m^{inj/out} = \tau_p\Gamma P_m^{inj/out}\lambda_m/(hc\cdot\eta_d\cdot V_w)$, where $\eta_d = 0.42$ represents optical efficiency and $h$ and $c$ are Planck's constant and speed of light, respectively. Finally, the last equations describe the dynamics of the phase difference ($\theta_m$) between the free-running (FR) and the injection-locked (IL) states for the modes $m$:

$$\frac{d\theta_m}{dt} = \frac{\alpha}{2}A_m + 2\pi\frac{c}{\lambda_m^2}\Delta\lambda_m - k_c\sqrt{\frac{S_m^{inj}}{S_m}}\sin\theta_m, \quad (4)$$

where $\alpha = 3$ is the linewidth enhancement factor.

To study the dynamical response of FP-LD on input optical Gaussian pulse-train with randomized peak powers, the system of coupled nonlinear differential equations (1-4) is solved using the adaptive step size Runge–Kutta method, to calculate $P_m^{out}$.

## III. FP-LD under Single Optical Injection

### A. Operating principle

Fig. 1(a) illustrates the operating principle of the proposed nonlinear element (NLE). A pulsed Gaussian optical signal from the master laser (ML), with a wavelength $\lambda_p$ and a fixed full width at half maximum (FWHM), is injected into the FP-LD via an optical circulator (OC). The FP-LD serves as an NLE, providing a peak power transfer function that maps the optical input $P_p^{inj}$ to the output signal $P_p^{out}$ [17]. The output spectrum of the FP-LD in the FR-like state, used in the experiment is shown in Fig. 1(b). Apart from the FP-LD's longitudinal modes and the ML's wavelength $\lambda_p$, the spectrum shows an additional wavelength $\lambda_c$ of the CW ML used in a dual-OI scheme. This section treats the case of zero optical power at $\lambda_c$. The output spectrum is filtered using a BPF centered around $\lambda_p$ to suppress the unnecessary wavelengths and enable NLEs cascadability in multilayer optical neural networks (NNs).

The difference in wavelengths between the ML and the nearest side-mode of the FP-LD, i.e., wavelength detuning $\Delta\lambda_p$, is shown to greatly influence the response of the FP-LD [15, 17]. The properties of the NAF are also subject to change with the variation in bias current $I$ of the FP-LD [17], as well as by adjustments of the BPF's central wavelength with respect to $\lambda_p$.

The transient (dynamic) behavior of the FP-LD can be analyzed using the carrier rate ($dn_w/dt$) versus $n_w$ plot, i.e., a phase plot, as demonstrated in [16, 17]. Fig. 1(c) illustrates stationary phase plots (colored lines) for several continuous-wave (CW) input optical powers injected into side mode $p = -3$ (minus sign represents a blue shift from the central mode $j = 0$), with a fixed wavelength detuning $\Delta\lambda_p = 0.206$ nm. For sufficiently low input power $P_p^{inj}$ (green line), there is a single stationary state ($dn_w/dt = 0$) corresponding to the FR-like regime, which is close to the carrier threshold concentration $n_{th}$ of FR laser (green circle marker). In this regime, FP-LD output power is high for the central mode, and low at $\lambda_p$. As $P_p^{inj}$ increases (red line), additional stable stationary state (IL$_p$) emerges at a lower carrier concentration than in the FR-like regime (red circle marker). For a range of $P_p^{inj}$, both stable stationary states can coexist (the region of static bistability), and the FP-LD state depends on the prehistory [16, 17]. For sufficiently high $P_p^{inj}$ (yellow line), the FR-like state is lost, and only IL$_p$ remains (yellow circle marker), characterized by the strong emission at $\lambda_p$ and the suppression of all other longitudinal modes.

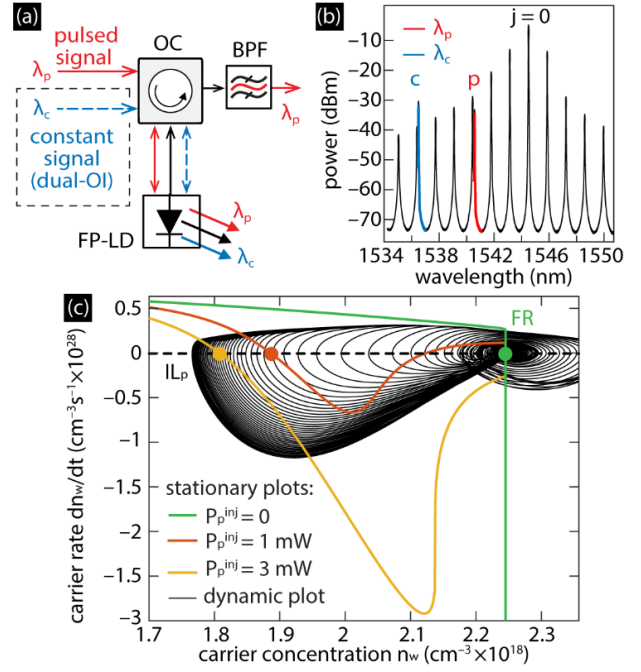

**Fig. 1**: (a) NLE operation, (b) Spectrum of the FP-LD in the FR state with optical injection $\lambda_p$ (and $\lambda_c$ in case of dual-OI) and (c) Phase plot of the FP-LD under optical injection $\lambda_p$ in the stationary (colored lines) and dynamic regime (thin black line).

The trajectory of the phase plot for the dynamic regime, under the injection of a Gaussian pulse train with a random sequence of peak optical powers, is shown as a thin black line. In response to the rising edge of a pulse, the dynamic trajectory attempts to reach the stationary IL$_p$ state. However, if the pulse is short enough (FWHM less than 1 ns) the IL$_p$ will rapidly cease to exist, resulting in return of the FP-LD to the FR state following the falling edge of the input pulse. This transition back to the FR state is carried out through strong relaxation oscillations, represented by a spiral loop of the dynamic trajectory. For sufficiently high input peak power and sufficient pulse duration, i.e., sufficient pulse energy, the dynamic trajectory reaches the corresponding stationary IL$_p$ state, providing high optical output power at $\lambda_p$. This behavior corresponds to the portion of the nonlinear transfer function (shown with a thin black line in Fig. 2(a)) above the activation threshold. Conversely, if the injected energy is low, the trajectory remains near the FR-like state, resulting in low $P_p^{out}$ and corresponding to the portion of the transfer function below the activation threshold.

Due to the rapid variations of carrier concentration within the FP-LD under pulsed optical injection, output pulses of NLE exhibit a rapid phase variation. This leads to the frequency chirping of the output signal (dotted line in Fig. 2(a)), resulting in the broadening of the FP-LD's output spectrum in the vicinity of the ML's wavelength. The simulation reveals that the output pulses below the activation threshold exhibit a frequency shift to the lower wavelengths (blue chirp), while the pulses above the threshold shift to higher wavelengths (red chirp), as illustrated in Fig. 2(a). Moreover, for a small wavelength detuning the blue-chirped region is practically inseparable from the nearest side mode $p$ of the FP-LD, which also exhibits spectral broadening and chirping due to pulses. Consequently, with the adjustment of the BPF, the ratio of the blue-chirped to the red-chirped signal in the output is altered.

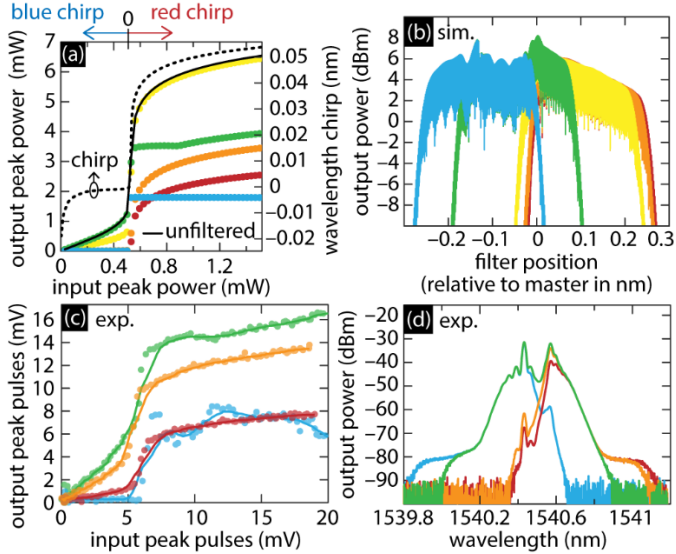

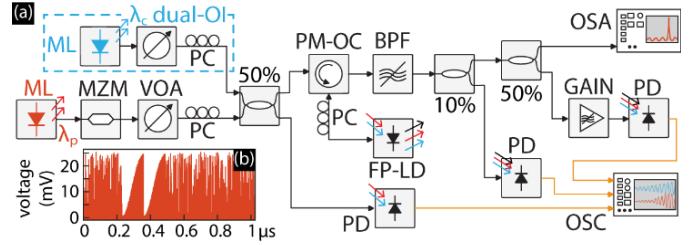

**Fig. 2.** The influence of BPF's position on the (a) simulated and (c) experimental NAF shapes and (b) simulated and (d) experimental FP-LD optical response spectra, for Gaussian pulses with FWHM = 362 ps at 167 MHz repetition rate. Dotted line and right axis in (a) correspond to simulated wavelength chirp.

**Fig. 3.** (a) Schematic of the experimental setup with an additional optional branch $\lambda_c$ for dual-OI (b) Input pulse patterns applied to the NLE (in this case, 50 ps wide Gaussians with the 0.5 GHz repetition rate).

This leads to the conclusion that the BPF's position with respect to the wavelength $\lambda_p$ provides an additional degree of freedom in the reconfiguration of the NAF's shape, as shown in Fig. 2(a). The corresponding simulated filtered spectra of the output signal are shown in Fig. 2(b). Based on the presented trends, we observe that the unfiltered NAF is decomposed into sections located within different spectral domains. By shifting the BPF into the red spectral region, the signal's blue-chirped spectral components are filtered out. This leads to the attenuation reduction of the linear part of the activation function below the threshold (yellow, red and orange solid markers in Fig. 2(a)). Alternatively, increasing the amount of the blue-chirped spectral components in the case of single-OI maintains the linear pre-threshold region, while rectifying the above threshold part of the activation (green and blue markers in Fig. 2). Finally, the low noise sigmoid-like activations are found in the red-chirped spectral region, with a small amount of added blue-chirp.

*B. Experimental demonstration*

Fig. 3(a) shows the experimental setup devised to investigate the FP-LD activations in the dual-OI scheme. Two tunable laser sources are used as the MLs with wavelengths $\lambda_p$ and $\lambda_c$. Their output powers and polarizations are controlled using electrically controlled variable optical attenuators (VOAs) and polarization controllers (PCs), respectively. The intensity of the $\lambda_p$ signal is modulated using a 40 Gb/s zero-chirp Mach-Zehnder modulator (MZM). Signals from both MLs are combined and injected into the FP-LD using a fast axis blocking polarization maintaining OC (PM-OC), while a third PC is employed between the PM-OC and the FP-LD. The sum of the injected pulsed and CW input signals is acquired by a PD connected to the oscilloscope (OSC). The FP-LD's output is further filtered using a variable BPF. The BPF adjusts the NLE's activation function prior to the monitoring section where the output signals are examined on the OSC and the optical spectrum analyzer (OSA). Polarizations of the input signals are adjusted to the slow axis of the PM-OC, while the third PC was used to compensate for the fiber birefringence between the PM-OC and the FP-LD chip within the laser's package, thus suppressing the strong polarization influence on the measurement. The GAIN block within the monitoring section is used to amplify the signal before detection and consists of an erbium-doped fiber amplifier (EDFA), additional VOA, and a second BPF. The unamplified NLE's output pulses are also monitored as a reference, to identify EDFA saturation effects throughout the measurements. All PDs within the setup have a bandwidth of 70 GHz, ensuring high-fidelity pulse acquisition.

The results of this paper are obtained for an FP-LD operating at a bias current of 14 mA (~1.5 times the laser's threshold current $I_{th}$), voltage of 1.02 V, and temperature of 27.2°C (near the operating temperature of the laser without the temperature control). After the BPF adjustment, its spectral width was set to 0.71 nm.

Fig. 3(b) shows representative traces of the input pulse patterns, acquired at the OSC. The patterns consisted of 2 linearly increasing pulse trains of 64 pulses each, followed by a sequence of 384 pulses with randomly generated peak powers. The pulses in the pattern are Gaussian pulses with varying width and repetition rate, generated using the MZM biased near its null point. The signals are averaged 8 times during acquisition, decreasing the noise accumulated along the line of amplification and measurement. Both the pulse widths and the repetition rates were varied in the experiments, examining the changes in the NLE's activations.

The experimental results verifying the influence of the BPF's position on the NAF shapes and spectrum are shown alongside the simulations of Fig. 2, in Fig. 2(c) and (d), respectively. The coloration between plots is carefully chosen to correlate the simulated and experimental trends that are alike, even though not all the filter parameters (e.g. filter width and position) were exactly the same. Although most of the trends predicted by simulations are confirmed by experiments, there is some discrepancy in results, i.e., the NAF denoted by green circles was always the highest in power in the experiments, while the NAF denoted by yellow circles was not experimentally reproduced. This can be explained by the imperfect parameter matching between the simulated and experimentally utilized (commercial) FP-LDs.

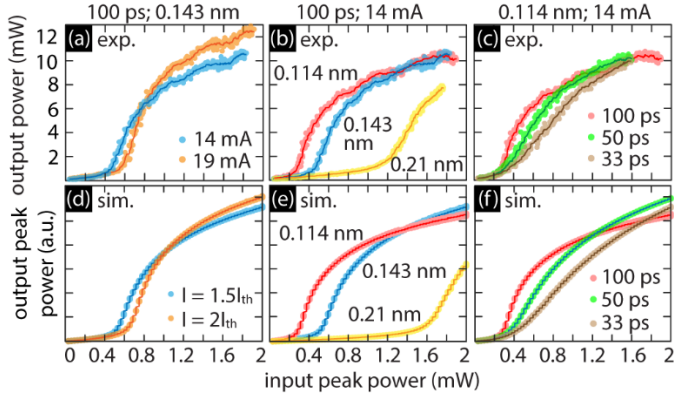

**Fig. 4.** Experimental and simulated results of NAF reconfigurability achieved by changing: (a) and (d) the bias current of the FP-LD, (b) and (e) detuning, (c) and (f) FWHM of input pulses for repetition rate 0.5 GHz and injection in $p = -3$.

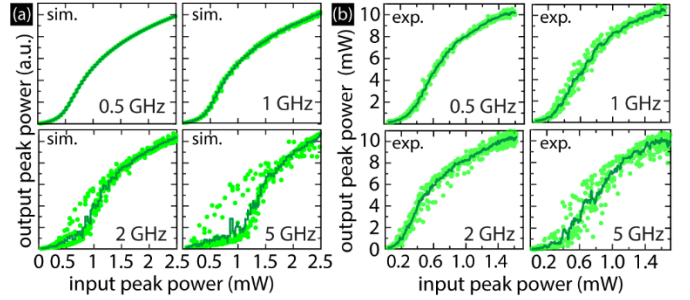

**Fig. 5.** The impact of increasing the repetition rate from 0.5 to 5 GHz on the (a) simulated and (b) experimentally determined NAFs for 50 ps pulses, mode $p = -3$ with a detuning of $\Delta\lambda_p = 0.119$ nm for the red-chirped part of the signal.

The reconfigurability of the nonlinear response for the filtered signal has been confirmed by the experimental results and numerical simulations shown in Fig. 4. The solid line in this and following figures of NAFs represents the result of a 10-point moving average of the experimental data depicted as discrete points. Figs. 4 (a) and (d) illustrate the impact of varying the bias current of the FP-LD, while (b) and (e) show the significance of detuning changes on the red-chirped filtered signal, respectively. The pulse width is 100 ps, and the repetition rate is 0.5 GHz. The bias current varies from 14 mA to 19 mA, corresponding to currents of $1.5I_{th}$ and $2I_{th}$. The effect of detuning variations from $\Delta\lambda_p = 0.119$ nm to 0.206 nm is more pronounced than the current variation, enabling the formation of sigmoid-like and PReLU-like nonlinearities [17, 15].

The response of the FP-LD to variations in the width of the input Gaussian pulses, for the red-chirped filtered signal, is shown in Figs. 4 (c) and (f). Increasing the pulse width produces a more prominent nonlinear response and steeper threshold region. For longer-duration pulses, the saturation region of the NAF occurs at lower input powers, consistent with the increased pulse energy.

An analysis of the repetition rate increase on the NAFs is shown on Fig. 5. The NAFs are obtained for a detuning of $\Delta\lambda_p = 0.119$ nm and pulses with a width of 50 ps, after filtering the red-chirped portion of the signal. The validity of the results obtained from the theoretical model (Fig. 5 (a)) is demonstrated by the excellent matching of the experimental data presented in Fig. 5 (b). As the repetition rate increases from 0.5 GHz to 5 GHz, the temporal separation between consecutive pulses decreases, preventing the FP-LD from fully recovering before the arrival of the next pulse leading to an increase in power variation in the laser's response. The recovery time of the FP-LD is determined by the effective carrier lifetime. In the single-OI process, the observed effect imposes a limitation on the operating speed, as further increase of repetition rate leads to severe performance degradation.

## IV. PERFORMANCE ENHANCEMENT THROUGH DUAL OPTICAL INJECTION

Speed enhancement is achieved through a dual-OI scheme that enables faster laser dynamics. This scheme involves the injection of an additional CW optical signal (dashed square marked "dual-OI" in Figures 1(a) and 3(a)) into $c = -6$, with a fixed wavelength detuning $\Delta\lambda_c = 0.103$ nm. The phase plot for the single-OI (c.f. Fig. 1(c)), which can exhibit bistability, is expanded under dual-OI to include additional stationary points, as shown in Fig. 6. This expansion allows for multistabililty and enables transitions involving an additional stationary state $IL_c$ and the aforementioned $IL_p$ stationary state [20, 21]. The injection of the additional optical signal, with a different detuning near the side mode $c$, modifies the stationary phase plot by introducing an additional "dip" in the phase trajectory. If the trajectory dip created solely by the CW signal injection (green line) is not deep enough to establish a stable stationary state, the dynamic phase plot trajectories are formed for transitions between the FR-like and $IL_p$ states. However, as illustrated in Fig. 6(a), the shape of the dynamic trajectory is altered, and the corresponding NAF's activation threshold is lowered (Fig. 7(a)). For sufficiently high CW input power $P_c$ (green line in Fig. 6(b)), however, an additional stable stationary point, $IL_c$, is established (green circle marker). This leads to injection locking of the FP-LD, independent of the pulsed signal injection. Consequently, the dynamic phase plot trajectories undergo a significant transformation: under pulsed injection, the trajectories now follow transitions between the $IL_c$ and $IL_p$ stable stationary states (thin black lines in Fig. 6(b)). As shown in the plot, the relaxation oscillations in this case are significantly suppressed, resulting in faster dynamics. Since these transitions involve smaller differences in carrier concentration, the peak output power and the activation threshold are both lower. This results in a substantial modification to the NAF shape compared to the cases of low CW input in dual- or single-OI. The reduced threshold and faster dynamics enhance the performance of the dual-OI scheme, particularly for high-speed optical applications.

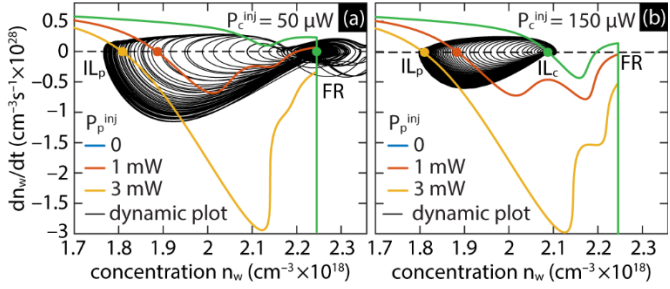

**Fig. 6.** Phase plot of the FP-LD under dual-OI $\lambda_p$ and $\lambda_c$ in the stationary and dynamic regimes for (a) low and (b) high CW input power.

The dual-OI scheme is accompanied by the same chirping effects observed within the single-OI, enabling further nonlinearity reconfiguration using the output BPF. However, the transition between the $IL_c$ and $IL_p$ states within the dual-OI paradigm is completely void of the presence of the FP-LD's noisy longitudinal modes, marking another key advantage of dual-OI nonlinearities. This enables the BPF to extract the blue-chirped region without the influence of the FP-LD's noisy longitudinal mode that significantly degrades the signal quality of the NAF output.

The NAFs obtained in a dual-OI concept, for different operating conditions are analyzed and presented in Figs. 7 and 8. The results of simulations and experiments for three characteristic CW laser power levels and filtering applied to the red-chirped signal, are presented in Fig. 7(a) and (b). The CW laser powers are chosen to represent single-OI, dual-OI starting from the FR-like state, and dual-OI starting from the $IL_c$ state, respectively. The increase in CW laser power does not affect the NAF shape significantly prior to achieving the $IL_c$ state (yellow and orange curves in Fig. 7(a) and (b)). The effect of adding the CW laser is thus limited to small shifts in threshold of the NAF toward lower input powers. In the context of artificial NN implementations, this phenomenon is an important achievement as it is equivalent to the addition of the optical bias [22]. The range of the NAF threshold shift is determined by the power required to achieve the stationary $IL_c$ state (120 µW in experiments). Further increase in CW laser power results in a transformation of the NAF shape, exhibiting a sudden change in the slope and saturation, as predicted by the phase plots.

The influence of the increase in speed on the obtained NAFs for the transition from the $IL_c$ state is analyzed through simulations and experiments, as shown in Fig. 7(c) and (d). The dual-OI paradigm shows significantly lower power fluctuation at 5 GHz speeds (red curve) compared to the results previously shown in the single-OI scheme (Fig. 5(b)). However, the sigmoid-like trend is observed to be less pronounced. Shifting the BPF from the red-chirped portion of the signal toward the ML's wavelength, changes the NAF profile from the sigmoid-like (red curve) to the inverted PReLU-like (blue curve), yielding a stronger nonlinearity. Since more of the signal passes through the BPF, higher output powers are achieved. For this BPF position, the repetition rate is increased up to 10 GHz for 25 ps pulses (green curve), while maintaining signal quality and slightly increasing the power fluctuation.

The reconfigurability of the $IL_c$ to $IL_p$ transition with respect to the detuning of the pulsed signal and the increase in CW laser power is shown in Fig. 8(a) and (b). The observed effect of detuning shifting the NAFs to higher powers, leads us to the conclusion that increasing the detuning $\Delta\lambda_p$ makes the activation functions more sigmoid-like. Furthermore, the output power range of the NAF decreases as the CW laser power increases, an effect explained by the reduction of the $IL_c$ state concentration on the phase plots. The smaller output power range in Fig. 8(a) and (b) compared to other results is a consequence of the measurement without an EDFA.

Fig. 8 (c) illustrates the relationship between the shape of the achieved NAF, filtered at $\lambda_p$, and the injected pulse width. The change in the shape of the $IL_c$ to $IL_p$ transition NAFs (filtered at $\lambda_p$) due to the fluctuation of pulse widths is shown in Fig. 8(c). The results are consistent with the assumption that higher energy pulses will provide lower threshold NAFs.

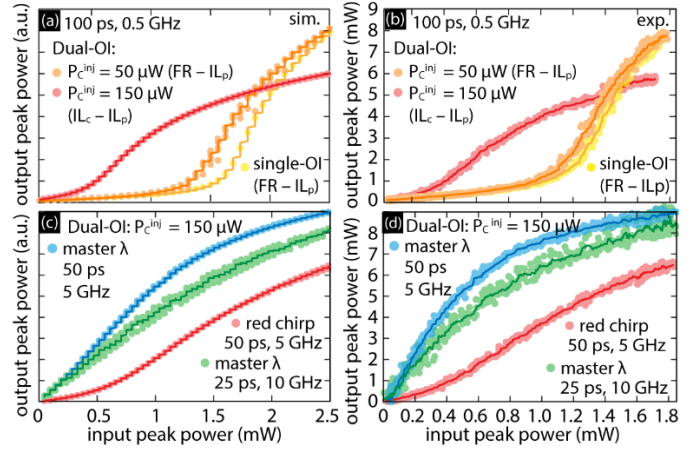

**Fig. 7.** Simulated (a, c) and experimental (b, d) NAFs obtained in the dual-OI scheme for $p = -3$, $\Delta\lambda_p = 0.279$ nm and $c = -6, \Delta\lambda_c = 0.143$ nm. (a) and (b) show the effect of the CW optical power increase, changing the NAFs from the yellow toward the red profile. (c) and (d) show the effect of the BPF shift from the red-chirp toward the ML's wavelength, shifting the NAF from the red to the blue profile. The increase in repetition rate is presented by the shift from the blue to the green NAF.

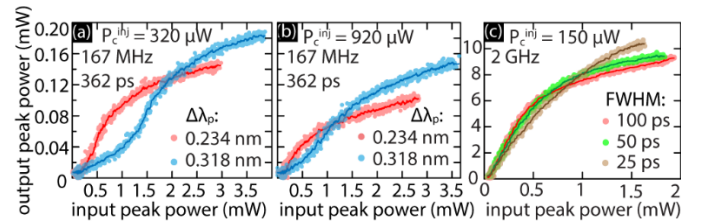

**Fig. 8.** Experimentally measured NAFs in the dual-OI scheme for the transition from the $IL_c$ to the $IL_p$ state under the influence of detuning for two CW powers (a) 320 µW and (b) 920 µW, and (c) for 25 to 100 ps pulse widths (detunings $\Delta\lambda_p = 0.206$ nm, $\Delta\lambda_c = 0.103$ nm).

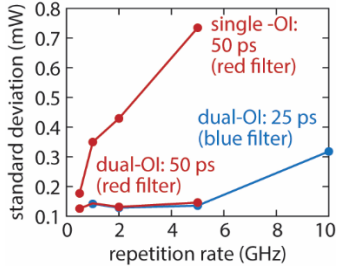

**Fig. 9.** The standard deviation of the NAFs with respect to the repetition rate.

To quantify the power fluctuation in experimental NAFs induced by the repetition rate increase, the standard deviation of data points from the moving average curve is calculated and presented in Fig. 9. In single-OI, as illustrated in Fig. 5, increasing the repetition rate causes a significant rise in power fluctuation, and consequently standard deviation. However, the results of dual-OI $IL_c$ to $IL_p$ transition show that the standard deviation remains nearly constant for repetition rates up to 5 GHz, regardless of pulse width and BPF position. Increasing the repetition rate to 10 GHz expectedly raises the standard deviation; however, it remains below the level observed in 1 GHz single-OI, emphasizing the enhanced high-speed optical capacity of the dual-OI scheme.

## V. ENERGY EFFICIENCY

A key performance indicator of an NLEs is energy consumption. Typically for NLEs, energy consumption is expressed in units of pJ per NOp, an approach usually assuming continuous incoming data at the highest achievable repetition rate. In passive devices, it is calculated by multiplying the pulse duration by the minimum optical peak power required for triggering non-linear phenomena. In active devices with continuous electrical power draw, it is typically calculated as the static consumed electrical power multiplied by the optical pulse duration, assuming that the highest achievable repetition rate is the inverse of the shortest attainable pulse duration. The energy of input optical pulses should be added to this calculation as well, although it is often negligible.

Table II puts in juxtaposition the experimentally achieved operating speed and energy consumption of the most prominent NAF demonstrations.

Table II: Comparison of this work and cutting-edge NAF implementations in terms of NAF type, speed and energy efficiency

| Ref. | Type | Speed [GHz] | Energy [pJ/NOp] |
|---|---|---|---|
| [4] | OEO | 10 | 37.5 |
| [11] | AO | 0.25 | 0.27-0.9 |
| [9, 10] | AO | 0.4-2.5 | 30 |
| [6] | AO | $10^{-4}$ | $7.4 \times 10^3$ |
| [8] | AO | 10 | 164 |
| **This work** | **AO** | **10** | **0.375-1.4** |

Based on the FP-LD's 14 mA current draw at an operating voltage of 1.02 V, its static power consumption is calculated to be 14 mW. Moreover, the activation threshold for the compared NLEs is approximated around 1 mW. Based on the outlined approach, and taking into account that our AO NAF uses 25 ps pulses at a repetition rate of 10 GHz, we achieve a low energy consumption of 1.4 pJ/NOp. We further estimate that the use of 25 ps pulses imposes a lower energy draw limit of just 375 fJ/NOp, calculated assuming successive pulses sent at a rate of 40 Gb/s with peak powers of 1 mW (corresponding to the threshold of the demonstrated activations). This is a record-breaking result for an active component implemented as an NLE. These results boast orders of magnitude improvements to previous demonstrations, putting our NLE next to the state-of-the-art passive nonlinear waveguides [11]. Even though our approach falls ~100 fJ/NOp short of the lowest power consumption demonstrated photonic NAF, it comes with some inherent advantages like higher reconfigurability, and no need for coherent pump-signal pulse pairs yielding easier cascadability in multilayer networks.

## VI. CONCLUSION

In this paper, we investigated AO nonlinear activation functions based on FP-LD bistability. The theoretical model, grounded in rate equations, has been experimentally validated in setups with one or two injected optical signals.

By characterizing the activation function in a single-OI scheme, we demonstrated complete reconfigurability of the activation function by tuning parameters such as bias current or detuning, as well as by adjusting the position of the BPF. This enables a wide range of activation functions, from sigmoid-like to PReLU-like. A repetition rate of up to 1 GHz was achieved in this scheme; however, further speed enhancement is quality-limited by the FP-LD recovery time. The challenge of achieving higher speeds was addressed in the dual-OI experiment by introducing an additional optical CW control signal.

The reconfigurability of the activation function depends on the same factors, accompanied by the effect of pulse width fluctuation. Notably, for pulse widths as low as 25 ps, a repetition rate of up to 10 GHz can be achieved while maintaining signal integrity and achieving energy consumption in the range of hundreds of fJ, making the realized activation functions strong candidates for applications in high-speed photonic neuromorphic computing.